\documentstyle[preprint,aps]{revtex}

\input{epsf.sty}

\def\ni{\noindent}
\def\ee{\vspace{3mm}}

\def\ket#1{\mbox{$| #1 \rangle$}}

\def\half{\textstyle{1 \over 2}}
\def\loud#1{\noindent{\bf #1 }}
\def\th{$^{\mbox{\scriptsize th}}\ $}
\def\thesize{\small}

\def\ox{\otimes}
\def\xor{\oplus}
\def\-{\leftarrow}
\def\01{\{0,1\}}
\def\x{\times}

\def\e{\varepsilon}
\def\d{\delta}
\def\px{\sigma_x}
\def\py{\sigma_y}
\def\pz{\sigma_z}
\def\SS{{\cal S}}
\def\G{\phi}

\begin{document}

\draft

\preprint{}

\title{Quantum Stabilizer Codes and Classical Linear Codes}

\author{Richard Cleve\thanks{{\tt cleve@cpsc.ucalgary.ca}}}
\address{\sl Department of Computer Science, University of Calgary,\\ 
Calgary, Alberta, Canada T2N 1N4 \\
{\rm and} \\
Institute for Theoretical Physics, University of California,\\
Santa Barbara, CA 93106}


\maketitle

\begin{abstract}
We show that within any quantum stabilizer code there lurks a classical 
binary linear code with similar error-correcting capabilities, 
thereby demonstrating new connections between quantum codes and 
classical codes.
Using this result---which applies to degenerate as well as nondegenerate 
codes---previously established necessary conditions for classical linear 
codes can be easily translated into necessary conditions for quantum 
stabilizer codes.
Examples of specific consequences are: for a quantum channel subject to 
a $\d$-fraction of errors, the best asymptotic capacity attainable by any 
stabilizer code cannot exceed $H(\half + \sqrt{2\d(1-2\d)})$; and, for the 
depolarizing channel with fidelity parameter $\d$, the best asymptotic 
capacity attainable by any stabilizer code cannot exceed $1-H(\d)$.
\end{abstract}


\section{Introduction}

The theory of error-correcting codes for classical information has been 
extensively studied for almost fifty years.
A fundamental question in coding theory concerns what {\em capacity} of 
information can be successfully transmitted through a noisy channel.
Call a code that maps $k$-bit inputs into $n$-bit codewords an $(n,k)$ 
code, and define its capacity to be $k/n$.
For any specific $n$-bit channel, let its {\em capacity} be the maximum 
capacity of all $(n,k)$ codes that successfully transmit information 
through it.
This capacity is often taken as an asymptotic limit as $n$ tends towards 
infinity.

We consider two basic kinds of $n$-bit noisy channels.
The first is one that flips any subset of up to $t$ bits of each 
codeword that passes through it.
In this model, transmission through the channel is considered successful if 
the $k$ bits of data can always be perfectly recovered.
Call a code that achieves this a {\em $t$-error-correcting code}.
Asymptotically, it is natural to take $t$ as some fixed fraction $\d$ of $n$, 
written as $\d n$ and understood to mean $\lfloor \d n \rfloor$.
A second model of a noisy channel is one where each bit of each codeword 
that passes through it is flipped independently with probability $\d$.
In this model, commonly referred to as the {\em binary symmetric channel\/}, 
there is no absolute bound on the number of possible errors that occur.
Therefore, a probabilistic definition of successful transmission is required. 
Call a code for which the probability of successful recovery for any $k$ 
bits of data is at least $1-\e$ an {\em $(\e,\d)$-error-correcting code}.

The subject of error-correcting codes for {\em quantum} information 
is much younger, developing within the past couple of years, though it 
has received considerable attention during this time 
\cite{Shor,CaSh,St1,LaMiPaZu,ShSm,BeDiSmWo,Gottesman,CRSS,St2,EkMa,KnLa,DiSh,%
CG,Ca2}.
Much of the above terminology extends naturally to quantum information by 
considering qubits instead of bits.
Call a quantum code that maps $k$-qubit data to $n$-qubit codewords 
an $((n,k))$ code.
We need to specify the behavior of noisy quantum channels.
A natural quantum analogue of the first model is to allow any $t$ qubits 
of each codeword that passes through it to be altered.
We can take this to mean: apply an arbitrary unitary transformation 
to all the qubits selected for alteration.
An apparently stronger definition allows the unitary transformation to 
also involve another set of qubits, representing an ``external environment'', 
thereby simulating the effect of ``decoherence''.
An apparently weaker definition limits the unitary operations to being among: 
$\px$, $\py$, and $\pz$, the standard Pauli spin matrices, and $I$, the unit 
matrix.
It turns out that, by reasoning similar to that in \cite{BeDiSmWo,EkMa,KnLa}, 
these three definitions of ``alter'' can be shown to be equivalent, in 
the sense that a code that is {\em $t$-error-correcting} with respect to 
the apparently weaker one will automatically be $t$-error-correcting with 
respect to the apparently stronger one.
A quantum analog of the binary symmetric channel is the {\em depolarizing 
channel}, where each bit of each codeword that passes through it is 
independently subjected to: $I$ with probability $1-\d$, and $\px$, $\py$, 
$\pz$ each with probability $\d/3$.
Call a code that achieves a fidelity of at least $1-\e$ on such a channel an 
{\em $(\e,\d)$-error-correcting code}.

It has previously been shown \cite{CaSh,St1,Ca2} how to take some special 
classical linear codes (binary and over $\mbox{\it GF}(4)$) with certain 
properties and transform them into quantum codes with error-correcting 
capabilities.
In the present paper, we show how to transform quantum codes 
with certain properties into classical codes with error-correcting 
capabilities.
We do not propose this as a means for constructing new classical codes; 
rather, this is a means for translating existing proofs 
of the {\em non}existence of certain classical codes into new 
proofs of the nonexistence of certain quantum codes.
Our specific results, which apply to the class of {\em stabilizer} quantum 
codes (defined in the next section), are:\ee

\loud{Theorem 1:}{\sl 
If there exists a $t$-error-correcting $((n,k))$ quantum stabilizer code 
then there exists a $t$-error-correcting $(n-1,k)$ classical binary linear 
code.}\ee

\loud{Theorem 2:}{\sl 
If there exists a $(\e,\d)$-error-correcting $((n,k))$ quantum stabilizer 
code then there exists a $(\e,\d)$-error-correcting $(n-1,k)$ classical 
binary linear code.}\ee

By these results, we can immediately assert that, when we restrict our 
attention to stabilizer codes, the classical upper bounds in 
\cite{MRRW,Shannon} apply.
In particular, when a quantum channel is subject to $\d n$ errors, 
the asymptotic capacity is bounded above by 
$H\left(\half + \sqrt{2\d(1-2\d)}\,\right)$, 
where $H$ is the binary entropy function defined as 
$H(p) = -p\log_2p-(1-p)\log_2(1-p)$.
In fact, a slightly stronger but more complicated upper bound is proven in 
\cite{MRRW}.
This stronger bound is plotted in FIG. 1.\ee

\centerline{\framebox{FIG. 1 here}}\ee

\ni This upper bound is stronger than the previously established $1-4\d$ 
bound in \cite{KnLa}, though the latter bound has the advantage that 
it applies to nonstabilizer codes as well.
It is noteworthy that all of the quantum codes proposed to date for the 
channels described above are stabilizer codes.
Other upper bounds exist for {\em nondegenerate} quantum codes 
(see \cite{EkMa} for a definition of nondegenerate). 
One is $1-H(\d)-p\log_23$, and is based on an analogue of the classical 
``sphere packing bound'' \cite{EkMa}, and another asserts that the 
asymptotic capacity is zero if $\d > 1/6$ \cite{Rains1}.
It remains an open question whether the $1-H(\d)-\d\log_23$ bound also 
applies to degenerate codes.
Very recently, it has been announced that the $\d > 1/6$ threshold bound 
{\em does} extend to nondegenerate codes, and this will be appear in a 
forthcoming paper \cite{Rains3}.
It is interesting to note that there exist some degenerate stabilizer codes 
that outperform all known nondegenerate codes on the depolarizing channel, 
for some values of $\d$ \cite{ShSm}.
The best lower bound for this channel that we are aware of is 
$1-H(2\d)-2\d\log_23$ \cite{CRSS}.

For the depolarizing channel with error probability $\d$, our results 
imply that, for stabilizer codes, the capacity is upper bounded by 
$1-H(\d)$, the bound for the classical binary symmetric channel 
\cite{Shannon}.\ee

\centerline{\framebox{FIG. 2 here}}\ee

\ni For some values of $\d$, this is stronger than the previously 
established upper bound of $1-4\d$ \cite{BeDiSmWo}, though the latter 
bound applies to nonstabilizer codes as well.
The best lower bound that we are aware of for this channel is 
$1-H(\d)-\d\log_23$ \cite{BeDiSmWo}, and a slightly larger value 
for some values of $\d$ \cite{ShSm}.

Should any improvements to the upper bounds in \cite{MRRW} for 
classical coding occur, they will automatically apply to quantum 
stabilizer codes.
Our results demonstrate interesting connections between quantum 
stabilizer codes and classical linear codes, and, for some instances 
of channels, yield stronger upper bounds than those that have appeared 
to date.

In Sections II and III, we provide a brief overview of quantum 
stabilizer codes and classical linear codes.
In Section IV, we describe how to construct a binary linear code from a 
quantum stabilizer code, and, in Section V, we show that this construction 
yields the error-correcting properties required for Theorems 1 and 2.

\section{Stabilizer Quantum Codes}

In \cite{Gottesman,CRSS} it has been shown that many quantum codes 
can be described in terms of {\em stabilizers}.
Define a stabilizer as a set of $n$-qubit unitary operators such that: 
each operator is a tensor product of $n$ matrices of the form 
$\px$, $\py$, $\pz$, and $I$, with a global phase factor of $\pm 1$; and,  
the set of operators is an abelian group.
The code that is defined by a stabilizer is the set of all $n$-qubit 
quantum states that are fixed points of each element of the stabilizer.
A stabilizer can be most easily described by a set of operators that 
generate it.
If one negates the phase factor of some of the generators, the resulting 
code will change, but will have identical characteristics to the original 
code.
Thus, one can always take the phase of each generator to be $+1$ without 
any loss of generality.

It is convenient to denote the generators of a stabilizer in the language 
of binary vector spaces, as in \cite{CRSS}.
Denote the generator 
$$G = U_{1} \otimes U_{2} \otimes \cdots \otimes U_{n}$$
as the $2n$ bit vector $(a|b)$, where, for $i \in \{1,\ldots,n\}$, 
$$a_i = 
\cases{1 & if $U_i =$ $\px$ or $\py$ \vspace*{-2mm}\cr
       0 & if $U_i =$ $I$ or $\pz$ \cr}$$
and 
$$b_i = 
\cases{1 & if $U_i =$ $\pz$ or $\py$ \vspace*{-2mm}\cr
       0 & if $U_i =$ $I$ or $\px$. \cr}$$
For example, $\px \ox I \ox \px \ox I \ox \pz \ox \py \ox \pz \ox \py$ 
is denoted as 
$(\, 1 \ 0 \ 1 \ 0 \ 0 \ 1 \ 0 \ 1 \,
| \, 0 \ 0 \ 0 \ 0 \ 1 \ 1 \ 1 \ 1 \,)$.
In this notation, the product of any two generators $(a|b)$ and 
$(a^{\prime}|b^{\prime})$ is equivalent (modulo a phase factor of $\pm1$) 
to $(a \xor a^{\prime}|b \xor b^{\prime})$, where $\xor$ denotes 
the bit-wise sum in modulo two arithmetic.
Also, $(a|b)$ and $(a^{\prime}|b^{\prime})$ commute if and only if 

\begin{equation}
(a \cdot b^{\prime}) \xor (a^{\prime} \cdot b) = 0,
\label{commute}
\end{equation}
where $\cdot$ denotes 
the inner product in modulo two arithmetic.

A stabilizer can then be written as an $m \x 2n$ matrix whose rows 
represent the generators.
For example, 
\begin{equation}
{\thesize
\left(\begin{array}{cccccccl|rccccccc}
1 & 1 & 1 & 1 & 1 & 1 & 1 & 1 \ & \ 0 & 0 & 0 & 0 & 0 & 0 & 0 & 0 
\vspace*{-2mm}\\
0 & 0 & 0 & 0 & 0 & 0 & 0 & 0 & 1 & 1 & 1 & 1 & 1 & 1 & 1 & 1 \vspace*{-2mm}\\
1 & 0 & 1 & 0 & 0 & 1 & 0 & 1 & 0 & 0 & 0 & 0 & 1 & 1 & 1 & 1 \vspace*{-2mm}\\
1 & 0 & 1 & 0 & 1 & 0 & 1 & 0 & 0 & 0 & 1 & 1 & 0 & 0 & 1 & 1 \vspace*{-2mm}\\
1 & 0 & 0 & 1 & 0 & 1 & 1 & 0 & 0 & 1 & 0 & 1 & 0 & 1 & 0 & 1 
\end{array}\right)
}
\label{eight}
\end{equation}
represents the eight generators of the stabilizer of a specific $((8,3))$ 
code that is 1-error-correcting (see \cite{Gottesman} for a detailed 
analysis of this code).

In general, if there are $n$ qubits and $m$ generators, we can encode 
$k = n-m$ data qubits (in the above example, $3 = 8 - 5$).
The error correcting capabilities of the code are related to commutativity 
relationships between the error operators and the generators 
\cite{Gottesman,CRSS}.

\section{Classical Binary Linear Codes}

An $(n,k)$ binary linear code is a $k$ dimensional subspace of 
$\01^n$ over modulo two arithmetic.
It is sufficient to specify a basis $M_1,\ldots,M_k$ for such a code.
Then the codeword for the $k$-bit string $x_1 \ldots x_k$ can be taken 
as the linear combination $x_1 M_1 \xor \cdots \xor x_k M_k$ 
(this mapping is a bijection between $\01^k$ and the code).
A natural way of specifying such a code is by an $k \x n$ {\em generator 
matrix}, whose rows are $M_1,\ldots,M_k$.
An example of such a code is 
\begin{equation}
{\thesize
\left(\begin{array}{ccccc}
1 & 0 & 1 & 1 & 0 \vspace*{-2mm} \\
0 & 1 & 0 & 1 & 1
\end{array}\right)}
\end{equation}
which is a $(5,2)$ code that is 1-error-correcting.

\section{Construction of Linear Codes from Stabilizer Codes}

Consider a quantum stabilizer code specified by a $2n \x m$ matrix $(X|Z)$.
We shall show how to construct the generator matrix of a classical 
binary linear code with similar error-correcting capabilities.

Our construction involves a transformation of the generator matrix into 
a useful standard form along the lines of that in \cite{CG}.
This conversion is accomplished by applying a series of basic 
transformations of the following two types, each of which leaves the 
error-correcting characteristics of the code unchanged.
The first is a {\em row addition}, where the $j$\th row is added to the 
$i$\th row, where $i\neq j$.
This corresponds to replacing the $i$\th generator with the product 
of the $i$\th and $j$\th generator, and setting its phase to $+1$.
The second is a {\em column transposition}, where $i$\th column is transposed 
with the $j$\th column in the submatrices $X$ and $Z$ simultaneously.
This corresponds to transposing the $i$\th qubit position with the 
$j$\th qubit position.

We begin by applying transformations of the above types to the matrix 
$(X|Z)$ in order to obtain 
\begin{equation}
\begin{array}{ll}
 & 
\hspace*{4mm} \overbrace{}^{s} \hspace*{1.6mm} 
\overbrace{}^{n-s} 
\hspace*{3.0mm} \overbrace{}^{n} \hspace*{3.5mm} 
\overbrace{}^{n-s}
\vspace*{-1mm}\\
\begin{array}{c}
\mbox{\scriptsize $s$} \ \{ \vspace*{-0.5mm}\\
\mbox{\scriptsize $r$} \ \{
\end{array}
&
\left(\begin{array}{c|c}
\begin{array}{c|r}
\, \mbox{\Large $I$}\,\, & \, \mbox{\Large $A$} \\ \hline 
\mbox{\LARGE 0}\,\, & \mbox{\LARGE 0} 
\end{array} \,
& \,
\begin{array}{c|c}
\mbox{\Large $E_1$} & \mbox{\Large $E_2$} \\ \hline
\mbox{\Large $E_3$} & \mbox{\Large $E_4$} 
\end{array}
\end{array}\right)
\end{array}
\label{stand1}
\end{equation}
where $s$ is the rank of the submatrix $X$, and $r = n-k-s$.
This is like performing Gaussian elimination on the $X$ submatrix.
Next, by performing row additions among the last $r$ generators 
and column transpositions among the last $n-s$ qubit positions, 
the matrix can be further converted to the form 
\begin{equation}
\begin{array}{ll}
 & 
\hspace*{4mm} \overbrace{}^{s} \hspace*{2mm} 
\overbrace{}^{t} \hspace*{2mm} \overbrace{}^{r_1} 
\hspace*{5.0mm} \overbrace{}^{s} \hspace*{2mm} 
\overbrace{}^{t} \hspace*{2mm} \overbrace{}^{r_1}
\vspace*{-1mm}\\
\begin{array}{c}
\mbox{\scriptsize $s$} \ \{ \vspace*{-0.5mm}\\
\mbox{\scriptsize $r_1$} \{ \vspace*{-0.5mm}\\
\mbox{\scriptsize $r_2$} \{ 
\end{array}
&
\left(\begin{array}{c|c}
\begin{array}{c|c|c}
\, \mbox{\Large $I$}\,\, & \mbox{\Large $A_1$} & \mbox{\Large $A_2$} \\ \hline 
\mbox{\LARGE 0}\,\, & \mbox{\LARGE 0} & \mbox{\LARGE 0} \\ \hline 
\mbox{\LARGE 0}\,\, & \mbox{\LARGE 0} & \mbox{\LARGE 0}
\end{array} \,
& \,
\begin{array}{c|c|c}
\mbox{\Large $B_1$} &\mbox{\Large $B_2$} & \mbox{\Large $B_3$} \\ \hline
\mbox{\Large $C_1$} & \mbox{\Large $C_2$} & \mbox{\Large $I$} \\ \hline
\mbox{\Large $D$}\,\, & \mbox{\LARGE 0} & \mbox{\LARGE 0}
\end{array}
\end{array}\right)
\end{array}
\label{stand2}
\end{equation}
where $r_1$ is the rank of $E_4$, $r_2 = r - r_1$, and $t = n - s - r_1$.
This is like performing Gaussian elimination on the submatrix $E_4$ of 
(\ref{stand1}).
Note that if $r_2 > 0$ and $D \neq 0$ and then one of the last 
$r_2$ generators would not commute with one of the first $s$ generators.
Therefore, we can set $r_2 = 0$, $r_1 = r$, and $t = k$.
Thus, the form (\ref{stand2}) becomes 
\begin{equation}
\begin{array}{ll}
 & 
\hspace*{4mm} \overbrace{}^{s} \hspace*{2mm} 
\overbrace{}^{k} \hspace*{2mm} \overbrace{}^{r} 
\hspace*{5.0mm} \overbrace{}^{s} \hspace*{2mm} 
\overbrace{}^{k} \hspace*{2mm} \overbrace{}^{r}
\vspace*{-1mm}\\
\begin{array}{c} 
\mbox{\scriptsize $s$} \ \{ \vspace*{-0.5mm}\\
\mbox{\scriptsize $r$} \ \{
\end{array}
&
\left(\begin{array}{c|c}
\begin{array}{c|c|c}
\, \mbox{\Large $I$}\,\, & \mbox{\Large $A_1$} & \mbox{\Large $A_2$} \\ \hline 
\mbox{\LARGE 0}\,\, & \mbox{\LARGE 0} & \mbox{\LARGE 0}
\end{array} \,
& \,
\begin{array}{c|c|c}
\mbox{\Large $B_1$} &\mbox{\Large $B_2$} & \mbox{\Large $B_3$} \\ \hline
\mbox{\Large $C_1$} & \mbox{\Large $C_2$} & \mbox{\Large $I$}
\end{array}
\end{array}\right)
\end{array}
\label{standard}
\end{equation}
where $s+k+r = n$.
Call any set of generators in this form (\ref{standard}) 
{\em in standard form}.

For a generator matrix in standard form, consider the classical binary 
linear code generated by the $k \x (n-r)$ matrix 
\begin{equation}
\begin{array}{ll}
 & 
\hspace*{4mm} \overbrace{}^{s} \hspace*{3mm} 
\overbrace{}^{k}
\vspace*{-1mm}\\
\begin{array}{c} 
\mbox{\scriptsize $k$} \ \{
\end{array}
&
\left(\begin{array}{c|c}
\mbox{\Large $A_1^T$} & \mbox{\ \Large $I$\ }
\end{array}\right).
\end{array}
\label{classical}
\end{equation}
We claim that this classical code has similar characteristics to the 
original quantum code.
Before stating this precisely, consider as an example the aforementioned 
$((8,3))$ code, that corrects one error, whose stabilizer was given by 
(\ref{eight}).
Converting it to standard form yields 
\begin{equation}
{\thesize
\left(\begin{array}{c|c}
\begin{array}{cccl|rcl|r}
1 & 0 & 0 & 0 \ & \ 1 & 1 & 1 \ & \ 0 \vspace*{-2mm} \\
0 & 1 & 0 & 0 & 1 & 1 & 0 & 1 \vspace*{-2mm} \\
0 & 0 & 1 & 0 & 1 & 0 & 1 & 1 \vspace*{-2mm} \\ 
0 & 0 & 0 & 1 & 0 & 1 & 1 & 1 \\ \hline 
0 & 0 & 0 & 0 & 0 & 0 & 0 & 0 
\end{array} \
& \
\begin{array}{cccl|rcl|r}
1 & 0 & 1 & 1 \ & \ 1 & 0 & 1 \ & \ 0 \vspace*{-2mm} \\
1 & 1 & 0 & 1 & 1 & 1 & 0 & 0 \vspace*{-2mm} \\
0 & 1 & 0 & 1 & 1 & 0 & 1 & 0 \vspace*{-2mm} \\
0 & 0 & 1 & 1 & 1 & 1 & 0 & 0 \\ \hline 
1 & 1 & 1 & 1 & 1 & 1 & 1 & 1 
\end{array}
\end{array}\right)
}
\end{equation}
which is an equivalent $((8,3))$ code.
The resulting binary linear code is 
\begin{equation}
{\thesize
\left(\begin{array}{cccl|rcc}
1 & 1 & 1 & 0 \ & \ 1 & 0 & 0 \vspace*{-2mm} \\
1 & 1 & 0 & 1 & 0 & 1 & 0 \vspace*{-2mm} \\
1 & 0 & 1 & 1 & 0 & 0 & 1 
\end{array}\right)}
\label{seven}
\end{equation}
which is a well-known $(7,3)$ code that corrects one error.
Thus, in this example, we obtain a classical code with slightly 
better capacity.
The reader may recall that previous constructions of $((7,1))$ quantum 
codes have been based on the same classical code (\ref{seven}) 
\cite{CaSh,St1}.
It should be noted that the present connection is quite different, as 
it involves an $((8,3))$ quantum code.

In the next section we shall show that, for any $t$-error-correcting 
$((n,k))$ quantum stabilizer code that is in standard form (\ref{standard}), 
the matrix (\ref{classical}) generates a $t$-error-correcting $(n-r,k)$ 
binary linear code.
Also, for the case of the depolarizing channel, we shall show that, for 
any $(\e,\d)$-error-correcting $((n,k))$ quantum stabilizer code in 
standard form (\ref{standard}), the matrix (\ref{classical}) generates 
an $(\e,\d)$-error-correcting $(n-r,k)$ binary linear code.

Furthermore, by applying operations of the types below, which also do not 
affect the error-correcting capabilities of the code \cite{DiSh}, we can 
guarantee the additional property that $r \ge 1$, which slightly sharpens 
the result.
The first operation is the {\em column switch}, in which the $i$\th column 
of submatrix $X$ is transposed with the $i$\th column of submatrix $Z$.
This corresponds to: in the $i$\th qubit position, changing each instance 
of a $\px$ in each generator to a $\pz$, and each instance of a $\pz$ to a 
$\px$ (while leaving each $I$ and $\py$ intact).
The second operation is the {\em column addition}, in which the $i$\th 
column of submatrix $Z$ is added to the $i$\th column of submatrix $X$.
This corresponds to: in the $i$\th qubit position, changing each instance 
of a $\py$ in each generator to a $\pz$, and each instance of a $\pz$ to a 
$\py$ (while leaving each $I$ and $\px$ intact).
See \cite{DiSh} for an explanation of why these two operations do not affect 
the characteristics of the code.

\section{Proof of Error-Correcting Properties of Construction}

In this section, we show that the constructions of the previous section 
satisfy the claimed error-correcting properties.

The classical code whose generator matrix is given by (\ref{classical}) 
consists of $2^k$ codewords in the space $\01^{n-r}$.
We shall construct an {\it isomorphism} between this code and a 
restricted version of the quantum code specified by (\ref{standard}).
The restricted version of the quantum code consists of $2^k$ codewords 
that are contained in a special set $\SS$ of $2^{n-r}$ distinct $n$-qubit 
states.
This set $\SS$ has the property that it is closed with respect to $\pz$ 
errors among the first $n-r$ qubit positions.
Intuitively, the effect of bit errors on classical codewords within the space 
$\01^{n-r}$ is equivalent to the effect of $\pz$ errors in the first $n-r$ 
qubit positions on quantum codewords within the space $\SS$.
Formally, the isomorphism that we shall construct is a mapping
$\G : \01^{n-r} \rightarrow \SS$, such that:
\begin{enumerate}
\item
$\G$ is bijective.
\item
For each $y_1 \ldots y_{n-r} \in \01^{n-r}$ that is a codeword of the 
classical code, $\G(y_1 \ldots y_{n-r})$ is a codeword of the quantum code.
\item
For each codeword $y_1 \ldots y_{n-r} \in \01^{n-r}$ of the classical 
code, and each error vector $e_1 \ldots e_{n-r} \in \01^{n-r}$, 
$$\G(y_1 \ldots y_{n-r} \xor e_1 \ldots e_{n-r}) 
= \pz^{e_1} \ox \cdots \ox \pz^{e_{n-r}} \ox 
\overbrace{I \ox \cdots \ox I\,}^r\, \G(y_1 \ldots y_{n-r}).$$
\end{enumerate}
The existence of such an isomorphism means that an error in the $i$\th bit 
of the classical code (for any $i \in \{1,\ldots,n-r\}$) corresponds to 
a $\pz$ error in the $i$\th qubit of the restricted version of the quantum 
code.
More precisely, if the quantum code can correct any $t$ errors then it can 
correct any $t$ $\pz$ errors among the first $n-r$ qubit positions, and 
then the following procedure for correcting any $t$ errors in the classical 
code exists. 
Given a codeword $y_1 \ldots y_{n-r}$ subjected to an error vector 
$e_1 \ldots e_{n-r}$ of weight bounded by $t$, first apply the mapping $\G$ 
to it.
By the second and third properties of $\G$, the result is 
$\G(y_1 \ldots y_{n-r})$ subjected to at most $t$ $\pz$ errors among the 
first $n-r$ qubit positions, which can therefore be corrected.
By the first property, $\G^{-1}$ can be applied to this corrected 
quantum codeword, yielding the correction of the original codeword.
Therefore, if we establish that there exists a $\G$ that satisfies the 
above three properties then the classical code specified by 
(\ref{classical}) must correct at least as many errors as the quantum code 
specified by (\ref{standard}).

For the case of the depolarizing channel with parameter $\d$, if the quantum 
code attains fidelity $1-\e$ then  it attains fidelity $1-\e$ for a channel 
that applies $\pz$ in each of the first $n-r$ qubit positions independently 
with probability $\d$ 
(in fact we may need to slightly modify the code by applying some column 
switch and column addition operations---defined in Section IV---along the 
lines of the ``twirling'' techniques explained in \cite{BeDiSmWo}).
Therefore, the corresponding classical code is correcting with probability 
at least $1-\e$ on a binary symmetric channel with parameter $\d$.
Thus, the existence of the above $\G$ also suffices for this noisy channel 
model.

In order to construct a bijection $\G$ with the above properties, 
we shall construct a useful basis for the quantum code.
We begin with the stabilizer specified by the matrix in standard form 
(\ref{standard}).
Call the operators corresponding to the respective rows of this matrix 
$G_1,\ldots,G_m$.
Define the additional operators $L_1,\ldots,L_k$ in terms of the matrix 
\begin{equation}
\begin{array}{ll}
 & 
\hspace*{4mm} \overbrace{}^{s} \hspace*{1.6mm} 
\overbrace{}^{k} \hspace*{2.4mm} \overbrace{}^{r} 
\hspace*{5.0mm} \overbrace{}^{s} \hspace*{2.2mm} 
\overbrace{}^{k} \hspace*{1.8mm} \overbrace{}^{r}
\vspace*{-1mm}\\
\begin{array}{c} 
\mbox{\scriptsize $k$} \ \{
\end{array}
&
\left(\begin{array}{c|c}
\begin{array}{c|c|c}
\, \mbox{\LARGE $0$}\,\, & \mbox{\ \Large $I$\ } & \mbox{\Large $C_2^T$}
\end{array} \,
& \,
\begin{array}{c|c|c}
\mbox{\,\Large $D$\,} &\mbox{\ \LARGE $0$\,} & \mbox{\ \LARGE $0$\,}
\end{array}
\end{array}\right),
\end{array}
\label{phase}
\end{equation}
where $D = B_2^T + C_2^T B_3^T$, 
and $N_1,\ldots,N_k$ in terms of the matrix 
\begin{equation}
\begin{array}{ll}
 & 
\hspace*{4mm} \overbrace{}^{s} \hspace*{2mm} 
\overbrace{}^{k} \hspace*{2mm} \overbrace{}^{r} 
\hspace*{5.0mm} \overbrace{}^{s} \hspace*{2.5mm} 
\overbrace{}^{k} \hspace*{1.5mm} \overbrace{}^{r}
\vspace*{-1mm}\\
\begin{array}{c} 
\mbox{\scriptsize $k$} \ \{
\end{array}
&
\left(\begin{array}{c|c}
\begin{array}{c|c|c}
\, \mbox{\LARGE $0$}\,\, & \mbox{\ \LARGE $0$\ } & \mbox{\ \LARGE $0$\,}
\end{array} \,
& \,
\begin{array}{c|c|c}
\mbox{\Large $A_1^T$} &\mbox{\ \Large $I$\ } & \mbox{\ \LARGE $0$\,}
\end{array}
\end{array}\right).
\end{array}
\label{bit}
\end{equation}
By considering (\ref{standard}), (\ref{phase}), (\ref{bit}), and 
recalling the criterion for commutativity (\ref{commute}), it is 
straightforward to verify that: 
\begin{itemize}
\item
$G_1,\ldots,G_m,L_1,\ldots,L_k$ is a set of $n$ independent 
commuting operators.
\item
$G_1,\ldots,G_m,N_1,\ldots,N_k$ is a set of $n$ independent 
commuting operators.
\item
Each $N_i$ and $L_j$ commute if $i \neq j$ and anticommute 
if $i=j$.
\end{itemize}

Using these properties, we can construct a basis 
$\{\ket{C_{x_1 \ldots\, x_k}} : x_1 \ldots x_k \in \01^k\}$ 
for the code with some useful structural features.
First, set $\ket{C_{0 \ldots 0}}$ to be the quantum state stabilized 
by $G_1,\ldots,G_m,L_1,\ldots,L_k$ (this state is unique up to a global 
phase factor).
In fact, 
\begin{equation}
\ket{C_{0 \ldots 0}} = 
{\textstyle{ 1 \over \sqrt{2^{s+k}} } }
(I + G_1) \cdots (I + G_s)
(I + L_1) \cdots (I + L_k)
\ket{\overbrace{0 \ldots 0}^n\,}
\label{c0}
\end{equation}
is a quantum state with this property.
Next, for each $x_1 \ldots x_k \in \01^k$, set 
$\ket{C_{x_1 \ldots\, x_k}} = 
N_1^{x_1} \cdots N_k^{x_k} \ket{C_{0 \ldots 0}}$.
Since $N_i$ commutes with $L_j$ if and only if $i=j$, 
$\ket{C_{x_1 \ldots\, x_k}}$ is in the +1-eigenspace of each 
$G_1,\ldots,G_m$ and the $(-1)^{x_1},\ldots,(-1)^{x_k}$ eigenspaces 
of $L_1,\ldots,L_k$, respectively.
Therefore, these states are an orthogonal basis for the quantum code.

Now, define the function $\G : \01^{n-r} \rightarrow \SS$ as 
\begin{equation}
\G(y_1 \ldots y_{n-r}) = 
\pz^{y_1} \ox \cdots \ox \pz^{y_{n-r}} \ox 
\overbrace{I \ox \cdots \ox I\,}^r\, 
\ket{C_{0 \ldots 0}},
\label{phi}
\end{equation}
for each $y_1 \ldots y_{n-r} \in \01^{n-r}$.
We shall show that $\G$ satisfies the three required properties.
By considering (\ref{standard}) and (\ref{phase}), the operator that 
applies $\pz$ to the $i$\th qubit (and $I$ to all other qubits) 
anticommutes with the $i$\th generator in the sequence 
$G_1,\ldots,G_s,L_1,\ldots,L_k$ and commutes with all others.
Therefore, $\G(y_1 \ldots y_{n-r})$ is in the 
$(-1)^{y_1},\ldots,(-1)^{y_{n-r}}$ 
eigenspaces of $G_1,\ldots,G_s,L_1,\ldots,L_k$, respectively (recall that 
$n-r = s+k$).
Thus, $\G(y_1 \ldots y_{n-r})$ is orthogonal for each distinct 
$y_1 \ldots y_{n-r}$.
This proves the first property, that $\G$ is a bijection.
Also, due to the close similarity between (\ref{classical}) and (\ref{bit}), 
\begin{equation}
\G(x_1 M_1 \xor \cdots \xor x_k M_k) 
= N_1^{x_1} \cdots N_1^{x_k} \ket{C_{0 \ldots 0}} 
=  \ket{C_{x_1 \ldots\, x_k}},
\label{similar}
\end{equation}
so the second property for $\SS$ holds.
Finally, the third property for $\SS$ holds because, using (\ref{phi}) and 
(\ref{similar}), 
\begin{eqnarray}
\G(y_1 \ldots y_{n-r} \xor e_1 \ldots e_{n-r}) 
& = & \pz^{y_1 \xor e_1} \ox \cdots \ox \pz^{y_{n-r} \xor e_{n-r}} \ox 
\overbrace{I \ox \cdots \ox I\,}^r\, 
\ket{C_{0 \ldots 0}} \nonumber\\
& = & \pz^{e_1} \ox \cdots \ox \pz^{e_{n-r}} \ox 
I \ox \cdots \ox I\, \G(y_1 \ldots y_{n-r}).
\end{eqnarray}
Thus, $\G$ satisfies the three required properties.

\acknowledgements

I am very grateful to 
Hans-Benjamin Braun for help in analyzing and plotting the functions in 
\cite{MRRW}, 
David DiVincenzo for several interesting discussions about quantum coding 
theory and comments about an earlier draft of this paper, 
Emanuel Knill for providing references to existing bounds for classical 
codes, and 
Juan Paz for interesting discussions about stabilizer representations 
of codes.
I am also grateful for the hospitality of the program on Quantum Computers 
and Quantum Coherence at the Institute for Theoretical Physics, University 
of California at Santa Barbara, where this work was completed.
This research was supported in part by NSERC of Canada and the U.S. 
National Science Foundation under Grant No.~PHY94-07194.

\begin{figure}
\epsfxsize=12cm
\leavevmode
\epsfbox{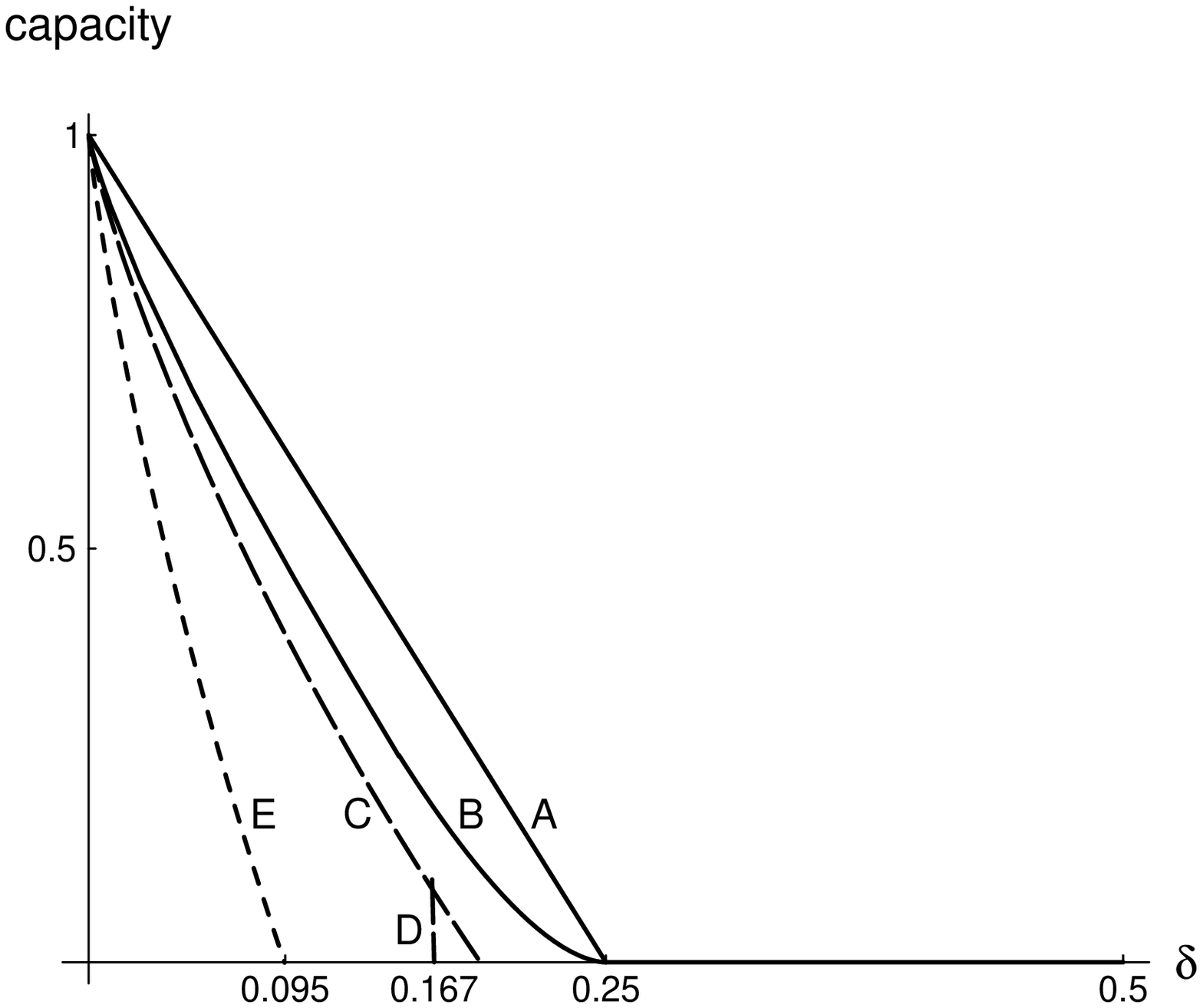}
\caption{Asymptotic upper bounds (solid lines), upper bounds for nondegenerate 
codes (broken lines), and lower bound (dashed line) 
for the capacity of a quantum channel with $\d$-bounded fraction of errors.
A: Linear upper bound in [11], 
B: Our new upper bound for stabilizer codes, based on the upper bounds for 
classical codes in [17] (see also [16]), 
C: The quantum ``sphere-packing'' upper bound for the case of nondegenerate 
codes in [10], 
D: The upper bound implied in [20] for nondegenerate codes, 
E: Lower bound in [14].}
\label{potential1}
\end{figure}

\begin{figure}
\epsfxsize=12cm
\leavevmode
\epsfbox{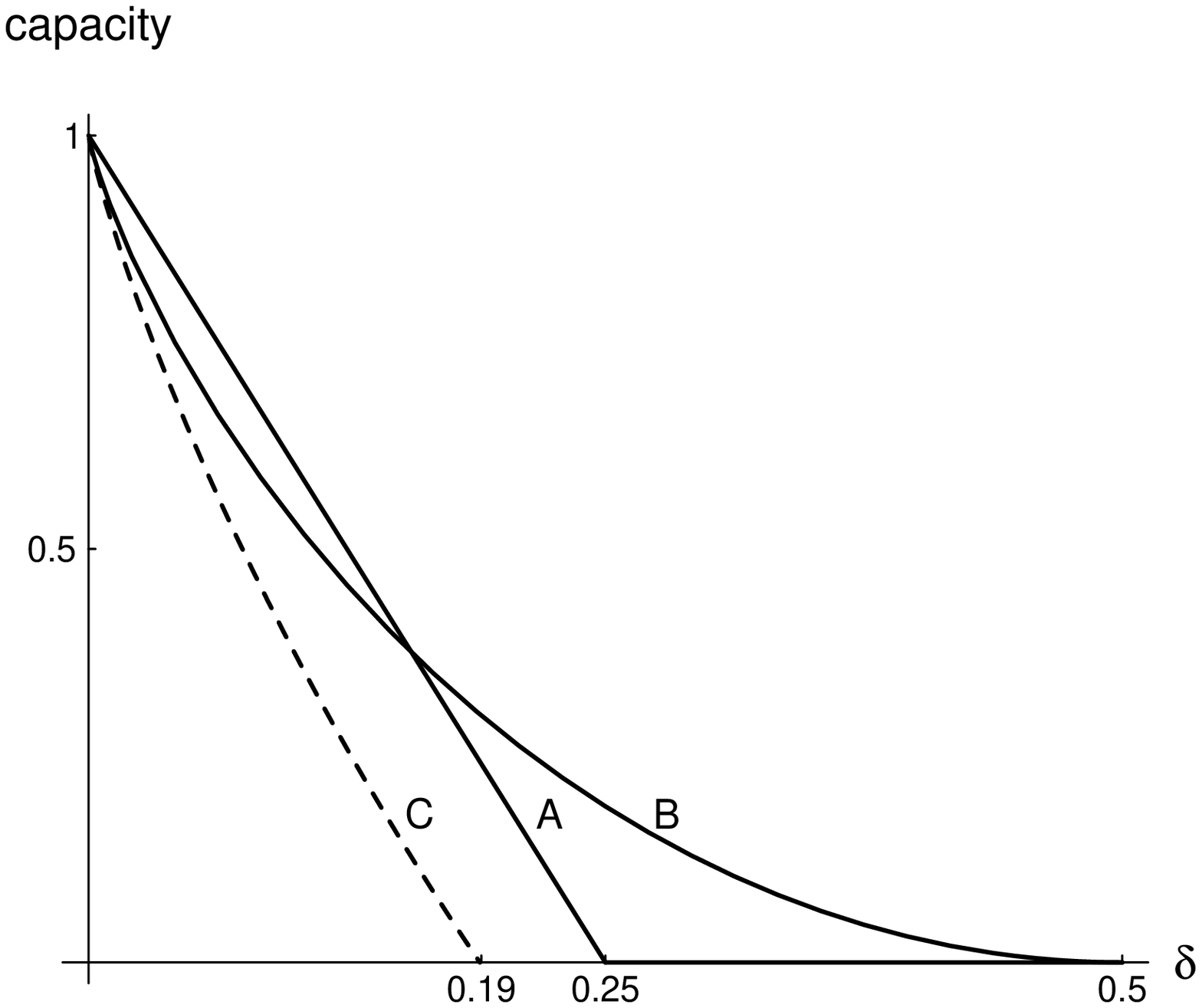}
\caption{Asymptotic upper bounds (solid lines) and lower bound (dashed line) 
for the capacity of the depolarizing quantum channel with probability 
parameter $\d$.
A: Linear upper bound in [6], 
B: Our new upper bound for stabilizer codes, based on the upper bounds for 
classical codes in [18] (see also [15]), 
C: Lower bound in [6] that matches the ``sphere-packing'' bound in [10], 
very slightly improved by [5].}
\label{potential2}
\end{figure}

\clearpage
\widetext


\begin{references}

\bibitem{Shor} P.W. Shor, ``Scheme for reducing decoherence in quantum 
memory,'' 
{\em Phys.\ Rev.\ A} Vol.~52, No.~4, pp.~2493--2496 (1995).

\bibitem{CaSh} A.R. Calderbank and P.W. Shor, 
``Good quantum error-correcting codes exist,''
{\em  Phys.\ Rev.\ A} Vol.~54, No.~2, pp.~1098--1105 (1996). 

\bibitem{St1} A.M. Steane, ``Error correcting codes in quantum theory,'' 
{\em  Phys.\ Rev.\ Lett.} Vol.~77, No.~5, pp~793--797 (1996). 

\bibitem{LaMiPaZu} R. Laflamme, C. Miquel, J.P. Paz, and W.H. Zurek, 
``Perfect quantum error correction code,'' 
{\em  Phys.\ Rev.\ Lett.} Vol.~77, No.~1, pp~198-- (1996). 

\bibitem{ShSm} P.W. Shor and J.A. Smolin, ``Quantum error-correcting codes 
need not completely reveal the error syndrome'', 
e-print {\tt quant-ph/9604006}.

\bibitem{BeDiSmWo} C.H. Bennett, D.P. DiVincenzo, J.A. Smolin, 
and W.K. Wooters, 
``Mixed state entanglement and quantum error correcting codes,'' 
{\em  Phys.\ Rev.\ A} Vol.~54, No.~5, pp.~3824--3851 (1996). 

\bibitem{Gottesman} D. Gottesman, ``A class of quantum error-correcting 
codes saturating the quantum Hamming bound,'' 
{\em  Phys.\ Rev.\ A} Vol.~54, No.~3, pp.~1862--1868 (1996).

\bibitem{CRSS} A.R. Calderbank, E.M. Rains, P.W. Shor, 
and N.J. Sloane, ``Quantum error correction and orthogonal geometry,'' 
e-print {\tt quant-ph/9605005}.

\bibitem{St2} A.M. Steane, ``Simple quantum error correcting codes,'' 
e-print {\tt quant-ph/9605021}.

\bibitem{EkMa} A. Ekert and C. Macchiavello, ``Error correction in quantum 
communication,'' 
{\em  Phys.\ Rev.\ Lett.} Vol.~77, No.~12, pp~2585--2588 (1996). 

\bibitem{KnLa} E. Knill and R. Laflamme, ``A theory of quantum 
error-correcting codes,'' 
e-print {\tt quant-ph/9604034}.

\bibitem{DiSh} D.P. DiVincenzo and P. Shor, 
``Fault-tolerant error correction with efficient quantum codes,'' 
{\em  Phys.\ Rev.\ Lett.} Vol.~77, No.~15, pp~3260--3263 (1996). 

\bibitem{CG} R. Cleve and D. Gottesman, ``Efficient computations of 
encodings for quantum error correction,''
e-print {\tt quant-ph/9607030}.

\bibitem{Ca2} A.R. Calderbank, E.M. Rains, P.W. Shor, 
and N.J. Sloane, ``Quantum error correction via codes over $GF(4)$,'' 
e-print {\tt quant-ph/9608006}.

\bibitem{MaSl} F.J. MacWilliams and N.J. Sloane, 
{\em The Theory of Error-Correcting Codes}, 
North Holland, Amsterdam, New York, Oxford, 1977.

\bibitem{vL} J.H. van Lint, {\em Introduction to Coding Theory}, 
Springer-Verlag, New York, 1982.

\bibitem{MRRW} R.J. McEliece, E.R. Rodemich, H.C. Rumsey, L.R. Welch, 
``New upper bounds on the rate of a code via the Delsarte-MacWilliams 
inequalities,'' 
{\em IEEE Trans.\ on Info.\ Theory}, Vol.~23, pp.\ 157--166, 1977.

\bibitem{Shannon} C.E. Shannon, ``A mathematical theory of communication,'' 
{\em Bell Sys.\ Tech.\ J.}, Vol.~27, pp.\ 379--423, 623--656, 1948. 

\bibitem{ShLa} P. Shor and R. Laflamme, 
``Quantum MacWilliams identities'' 
e-print {\tt quant-ph/9610040}.

\bibitem{Rains1} E. Rains, 
``Quantum shadow enumerators,'' 
e-print {\tt quant-ph/9611001}.

\bibitem{Rains2} E. Rains, 
``Quantum weight enumerators,'' 
e-print {\tt quant-ph/9612015}.

\bibitem{Rains3} E. Rains, private communication, December 1996.

\end{references}
\end{document}